\documentclass[12pt]{article}

\usepackage{epsf}
\usepackage[dvips]{graphicx}

\usepackage{latexsym}

\def\bc{\begin{center}}
\def\ec{\end{center}}
\def\beq{\begin{equation}}
\def\eeq{\end{equation}}

\def\sig {\sigma}

\def\hs#1{\hspace*{#1cm}}

\def\av#1#2{\langle {#1} \rangle^{}_{#2}}
\def\ave#1{\langle {#1} \rangle}

\def\F{{n^{}_F}}
\def\B{{n^{}_B}}

\def\pF{{p^{}_{{\rm t}F}}}
\def\pB{{p^{}_{{\rm t}B}}}

\def\DF{\Delta y_F^{}}
\def\DB{\Delta y_B^{}}

\def\bfs{{\bf s}}
\def\bfb{{\bf b}}

\def\j#1#2#3#4{{#1} {\bfseries  #2}, #4 (#3)}

\def\NP{Nucl. Phys.}
\def\PL{Phys. Lett.}
\def\PRL{Phys. Rev. Lett.}
\def\PR{Phys. Rev.}

\def\PRep{Phys. Rep.}

\def\EPJ{Eur. Phys. J.}

\begin{document}

\begin{center}
{\bfseries MULTIPLICITY AND {\Large $p_t$} CORRELATIONS\\
 IN RELATIVISTIC NUCLEAR COLLISIONS}

\vskip 5mm

R.S. Kolevatov and V.V. Vechernin$^{\dag}$

\vskip 5mm

{\small
{\it
V.A.Fock Institute of Physics, St.Petersburg State University
}
\\
$\dag$ {\it
E-mail: vecherni@pink.phys.spbu.ru
}}
\end{center}

\vskip 5mm

\begin{center}
\begin{minipage}{150mm}
\centerline{\bf Abstract}

The theoretical description of the correlations between observables
in two separated rapidity intervals for AA-interactions at high energies
is presented.
In the case with the real nucleon distribution density of colliding nuclei
the MC calculations of the long-range correlation functions
at different values of impact parameter are done.

For $n$--$n$ and $p_t$--$n$ correlations it is shown that
the impact parameter fluctuations at a level of a few fermi,
unavoidable in the experiment, significantly
change the magnitude of correlation coefficients.
The rise of $p_t$--$n$ and especially $p_t$--$p_t$ correlation
coefficients is found when one passes from SPS to RHIC and LHC
energies.
\end{minipage}
\end{center}

\vskip 10mm

\section{String fusion model (SFM)}

The colour string model \cite{Capella1,Kaidalov}
originating from Gribov-Regge approach
is being widely applied
for the description of the soft part of the multiparticle production
in hadronic and nuclear interactions at high energies.
In this model
at first stage of hadronic interaction
the formation of the extended objects -
the quark-gluon strings - takes place.
At second stage
the ha\-d\-ro\-ni\-za\-tion of these strings produces the observed hadrons.
In the original
version the strings evolve independently and the observed spectra are just
the sum of individual string spectra.
However in the case of nuclear collision,
with growing energy and atomic number of colliding nuclei,
the number of strings grows and
one has to take into account the interaction between them.

One of possible approaches to the problem
is the colour string fusion model \cite{BP1}.
The model is based on a simple observation
that due to final transverse dimensions of strings
they inevitably have to start to overlap
with the rise of their density in transverse plane.
At that the interaction of string colour fields takes place,
which changes the process of their fragmentation into hadrons
as compared with the fragmentation of independent strings.
So we have one more interesting nonlinear phenomenon
in nuclear interactions at high energies - the field of physics
the investigations in which were initiated by pioneer works
of academician A.M.~Baldin \cite{Baldin}.

%%%%%%%%%%%%%%%
It was shown \cite{BP1,BP00EPJC,BP00PRL} that the string fusion phenomenon considerably damps
the charged particle multiplicity and simultaneously
increase their mean $p_t$ value
as compared with the case of independent strings.
In accordance with a general Schwinger idea \cite{Schwinger51}
and the following papers \cite{Biro,Bialas} (colour ropes model)
two possible versions of string fusion mechanism were suggested.

The first version \cite{BP00EPJC} of the model assumes that
the colour fields are summing up only locally
in the area of overlaps of strings in the transverse plane.
So we will refer to this case as a \textit{local} fusion
or \textit{overlaps}. In this case one has
\beq
\av{n}{k}=\mu_0\frac{S_k}{\sig_0}\sqrt{k} \hs2
\av{p^2_t}{k}=p^2\sqrt{k}
\label{local}
\eeq
Here $\av{n}{k}$ is the average multiplicity of charged particles
originated from the area $S_k$, where $k$ strings are overlapping,
and $\av{p^2_t}{k}$ is the same for their squared transverse momentum.
The $\mu_0$ and $p^2$ are the average multiplicity
and squared transverse momentum of charged particles produced
from a decay of one single string,
and $\sig_0$ is its transverse area.

In the second version \cite{BMP02PRC} of the model one assumes that
the colour fields are summing up globally
- over total area of each cluster in the transverse plane -
into one average colour field.
This case corresponds to the summing of the source colour charges.
We will refer to this case as a \textit{global} fusion
or \textit{clusters}. In this case we have
\beq
\av{n}{cl}=\mu_0\frac{S_{cl}}{\sig_0}\sqrt{k_{cl}} \hs1
\av{p^2_t}{cl}=p^2\sqrt{k_{cl}} \hs1
k_{cl}=\frac{ N^{str}_{cl} \sig_0 }{ S_{cl} }
\label{global}
\eeq
Here $\av{n}{cl}$ is the average multiplicity of charged particles
originated from the cluster of the area $S_{cl}$
and $\av{p^2_t}{cl}$ is the same for their squared transverse momentum.
The $N^{str}_{cl}$ is the number of strings forming the cluster.

Note that in two limit cases both versions give the same results.

For $N$ non-overlapping strings we have in the \textit{local} version:
$k=1$, $S_1=N\sig_0$, $\ave{n}=\av{n}{1}=N\mu_0$ and $\ave{p^2_t}=\av{p^2_t}{1}=p^2$.
In the \textit{global} version in this case we have $N$ clusters
each formed by only one string, so $k_{cl}=1$, $\ave{n}=N\av{n}{cl}=N\mu_0$
and $\ave{p^2_t}=\av{p^2_t}{cl}=p^2$.

For $N$ totally overlapped strings we have in the \textit{local} version:
$k=N$, $S_N=\sig_0$, $\ave{n}=\av{n}{N}=\sqrt{N}\,\mu_0$
and $\ave{p^2_t}=\av{p^2_t}{N}=\sqrt{N}\,p^2$.
In the \textit{global} version in this case we have one cluster of the area $S_{cl}=\sig_0$
formed by $N$ string, so $k_{cl}=N$, $\ave{n}=\av{n}{cl}=\sqrt{N}\,\mu_0$
and $\ave{p^2_t}=\av{p^2_t}{cl}=\sqrt{N}\,p^2$.

So in both versions of the model when we pass from $N$ non-overlapping strings
to $N$ totally overlapped strings the average multiplicity decreases from
$\ave{n}=N\mu_0$ to $\ave{n}=\sqrt{N}\,\mu_0$ and the mean $p^2_t$
increases from $\ave{p^2_t}=p^2$ to $\ave{p^2_t}=\sqrt{N}\,p^2$.

%%%%%%%cells==============
\section{Cellular analog of SFM}

To simplify calculations in the case of real nucleus-nucleus collisions
a simple cellular model
originating from the string fusion model was proposed \cite{Vestn}.
In the framework of the cellular analog
along with the calculation simplifications
the asymptotics of correlation coefficients at large and small
string densities can be found analytically
in the idealized case with the homogeneous string distribution,
which enables to use these asymptotics later for the control of the
Monte-Carlo (MC) algorithms.

Two versions of the cellular model as the original SFM
can be formulated - with local and global string fusion.
In this model we divide all transverse (impact
parameter) plane into sells of order of the transverse string size $\sig_0$.

In the version with {\it local} fusion
the assumption of the model is
that if the number of strings
belonging to the $ij$-th cell is $k_{ij}$, then they form
higher colour string, which emits in average $\mu _0\sqrt{k_{ij}}$ particles
with mean $p_t^2$ equal to $p^2\sqrt{k_{ij}}$ (compare with (\ref{local})).
Note that zero "occupation numbers" $k_{ij}=0$ are also admitted.

In the version with {\it global} fusion at first
we define the neighbour cells as the cells with a common link.
Then we define the cluster as the set of neighbour cells
with non zero occupation numbers $k_{ij} \neq 0$.
After that we can apply the same formulae of the global fusion (\ref{global})
as in the original SFM,
where $N^{str}_{cl}$ is the number of strings in the cluster
and $S_{cl}/\sig_0$ is the number of cells in the cluster.

From event to event the number of strings $k_{ij}$ in the $ij$-th cell
will fluctuate around some average value - $\overline{k}_{ij}$.
Clear that in the case of real nuclear collisions
these average values $\overline{k}_{ij}$ will be different
for different cells. They will depend on the position ($\bfs_{ij}$)
of the $ij$-th cell in the impact parameter plane
(${\bf s}$ is two dimensional vector in the transverse plane).
In the case of nucleus-nucleus $AB$-collision
at some fixed value of impact parameter $\bfb$
one can find
this \textit{average} local density of primary strings $\overline{k}_{ij}$
in the point $\bfs_{ij}$
using nuclear profile functions $T_A(\bfs_{ij}+\bfb/2)$ and $T_B(\bfs_{ij}-\bfb/2)$.

In MC approach
knowing the $\overline{k}_{ij}$ one can generate some configuration $C\equiv\{k_{ij}\}$.
To get the physical answer for one given event (configuration $C$)
we have to sum the contributions
from different cells in accordance with \textit{local} or \textit{global} algorithm
(see above),
which corresponds to the integration over ${\bf s}$ in transverse plane.
Then we have to sum over events (over different configurations $C$).
Note that as the event-by-event fluctuations of the impact parameter at a level of a few fermi
are inevitable in the experiment one has to include the impact
parameter $b$ into definition of configuration $C\equiv\{b,k_{ij}\}$.

\section{Long-range correlations}

The idea \cite{BP00EPJC,BP00PRL,ABP94PRL} to use
the study of long-range correlations in nuclear collisions
for observation of the colour string fusion phenomenon based
on the consideration that the quark-gluon string is an extended object
which fragmentation gives the contribution to wide rapidity range.
This can be an origin of the long-range correlations in rapidity space
between observables in two different and separated rapidity intervals.
Usually in an experiment
they choose these two separated rapidity intervals
in different hemispheres of the emission of secondary particles
one in the forward and another in the backward
in the center mass system.
So sometimes these long-range rapidity correlations are referred as
the forward-backward correlations (FBC).

In principle one can study three types of such long-range correlations:
\begin{description}
  \item[$n$-$n$] - the correlation between multiplicities of charged particles
in these rapidity intervals,
  \item[$p_t$-$p_t$] - the correlation between transverse momenta
in these intervals and
  \item[$p_t$-$n$] - the correlation between the transverse momentum
in one rapidity interval and the multiplicity of charged particles
in another interval.
\end{description}
Usually to describe these correlations numerically
one studies the average value $\av{B}{F}$
of one dynamical variable $B$ in the backward rapidity window $\DB$,
as a function of another dynamical variable $F$
in the forward rapidity window $\DF$.
Here $\av{...}{F}$ denotes averaging over events
having a fixed value of the variable $F$ in the forward rapidity window.
The $\ave{...}$ denotes averaging over all events.
So we find the correlation function $\av{B}{F}=f(F)$.

It's naturally then to define the correlation coefficient
as the response of $\av{B}{F}$ on the variations of the variable
$F$ in the vicinity of its average value $\ave{F}$.
At that useful also to go to the relative variables,
i.e. to measure a deviation of $F$ from its average value $\ave{F}$
in units of $\ave{F}$, and the same for $B$.
So it's reasonable to define a correlation coefficient $b^{}_{B-F}$
for correlation between observables $B$ and $F$
in backward and forward rapidity windows in the following way:
\beq
b^{}_{B-F}\equiv\frac{\ave{F}}{\ave{B}} \left.\frac{d\av{B}{F}}{dF}\right|_{F=\ave{F}}
\label{bB-F}
\eeq

As the dynamical variables we use the multiplicity of charged particles ($n$),
produced in the given event in the given rapidity window,
and the \textit{event(!)} mean value of their transverse momentum ($p_t$),
i.e. the sum of the transverse momentum magnitudes of all charged particles,
produced in the given event in the given rapidity window ($\Delta y$),
divided by the number of these particles ($n$):
\beq
p_t \equiv \frac{1}{n} \sum_{i=1}^{n} |{\bf p}_{ti}|,
\hs1 {\rm where} \hs1 y_i\in\Delta y;
%\hs1 {\rm for}
\hs1 i=1,...,n.
\label{pt}
\eeq
So we can define three correlation coefficients:
$$
b^{}_{n-n}\equiv\frac{\ave{\F}}{\ave{\B}}\left.\frac{d\av{\B}{\F}}{d\F}
\right|_{\F=\ave{\F}}
%\label{bn-n}
\hs2
b^{}_{p_t-p_t}\equiv\frac{\ave{\pF}}{\ave{\pB}}\left.\frac {d\av{\pB}{\pF}} {d\pF}
\right|_{\pF=\ave{\pF}}
%\label{bpt-pt}
$$
\beq
b^{}_{p_t-n}\equiv\frac{\ave{\F}}{\ave{\pB}}\left.\frac {d\av{\pB}{\F}} {d\F}
\right|_{\F=\ave{\F}}
\label{bpt-n}
\eeq
Here $\B$, $\F$ are the multiplicities and $\pB$, $\pF$  are the
\textit{event} (\ref{pt}) mean transverse momentum of the charged particles,
produced in the given event
correspondingly in the backward ($\DB$) and forward ($\DF$) rapidity windows.

%%%%%%%%%%%%%||||||||||||||||||||||||
%%%%%% Fig.1 %%%%%%%%%%%%%%%%%%%%%
\begin{figure}[t]
\epsfysize=140mm
 \centerline{
\epsfbox{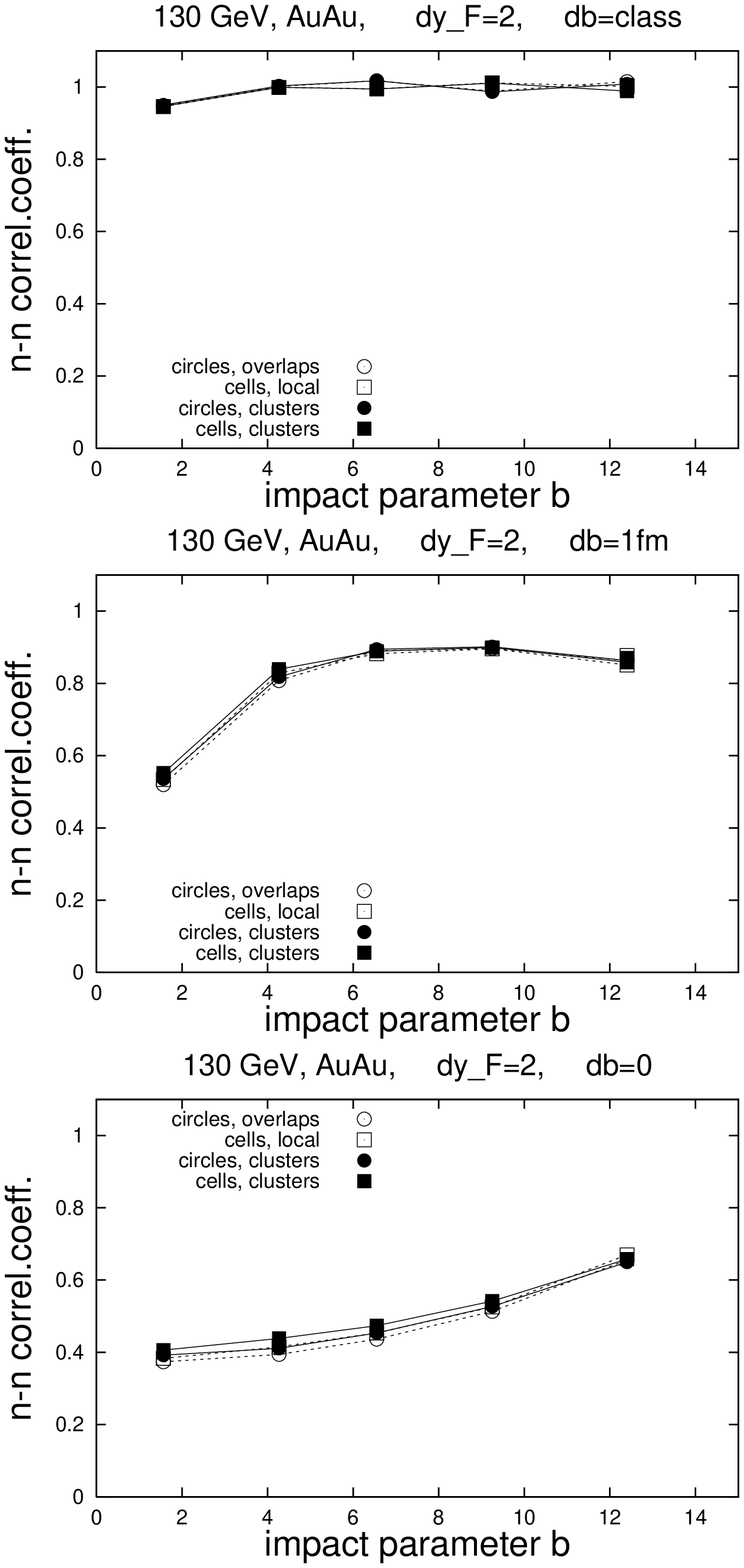}
 }
 \caption{The $b^{}_{n-n}$ correlation coefficient for
AuAu collisions at $\sqrt{s}=130$ GeV as a function of the impact parameter $b$
for tree choices of impact parameter window $db$ (see text).}
\end{figure}
%%%%%% Fig.2 %%%%%%%%%%%%%%%%%%%%%
\begin{figure}[t]
 \epsfysize=140mm
 \centerline{
 \epsfbox{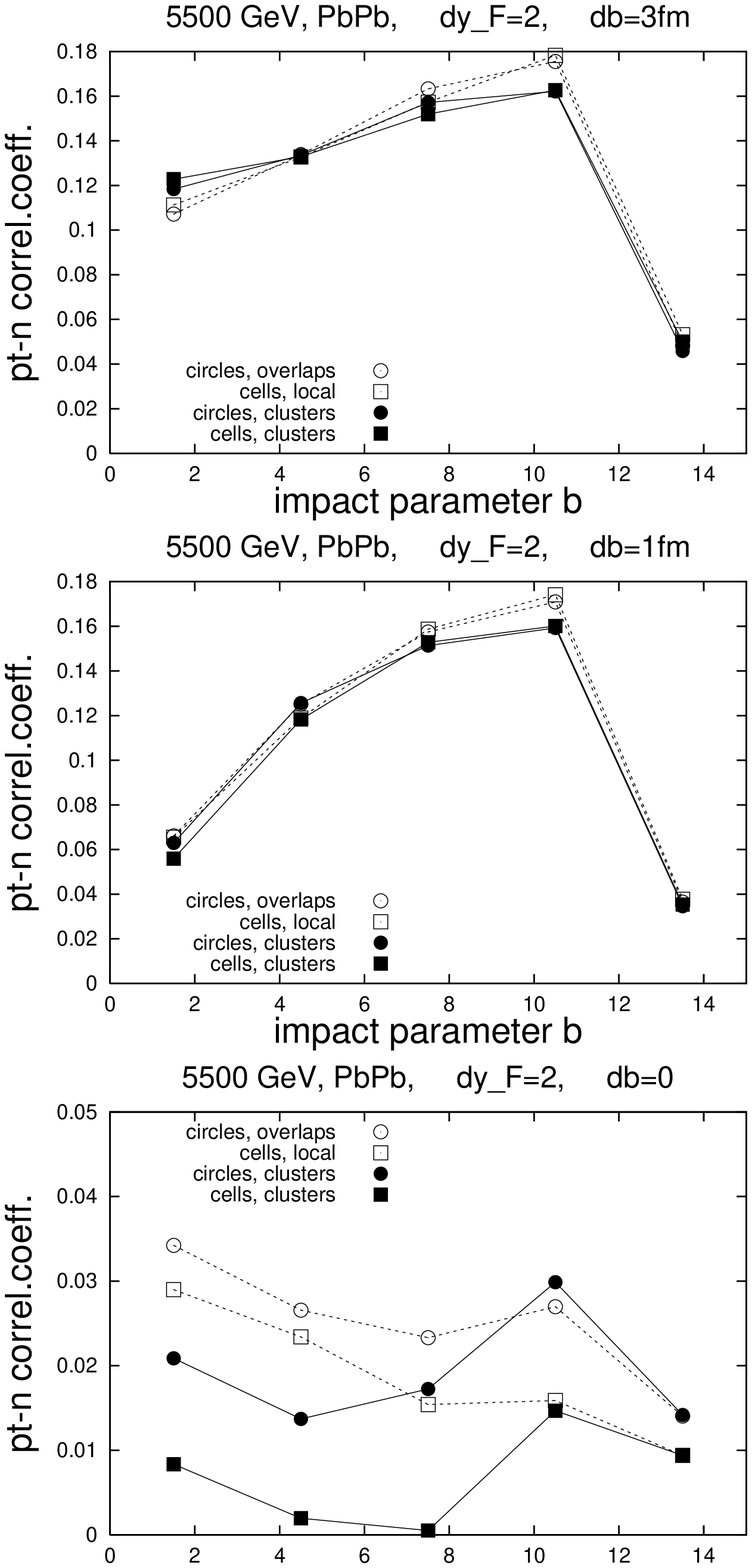}
 }
 \caption{The same as in Fig.1 but for the $b^{}_{p_t-n}$ correlation coefficient
 for PbPb collisions at $\sqrt{s}=5500$ GeV.}
\end{figure}
%%%%%% Fig.3 %%%%%%%%%%%%%%%%%%%%%
\begin{figure}[t]
 \epsfysize=140mm
 \centerline{
 \epsfbox{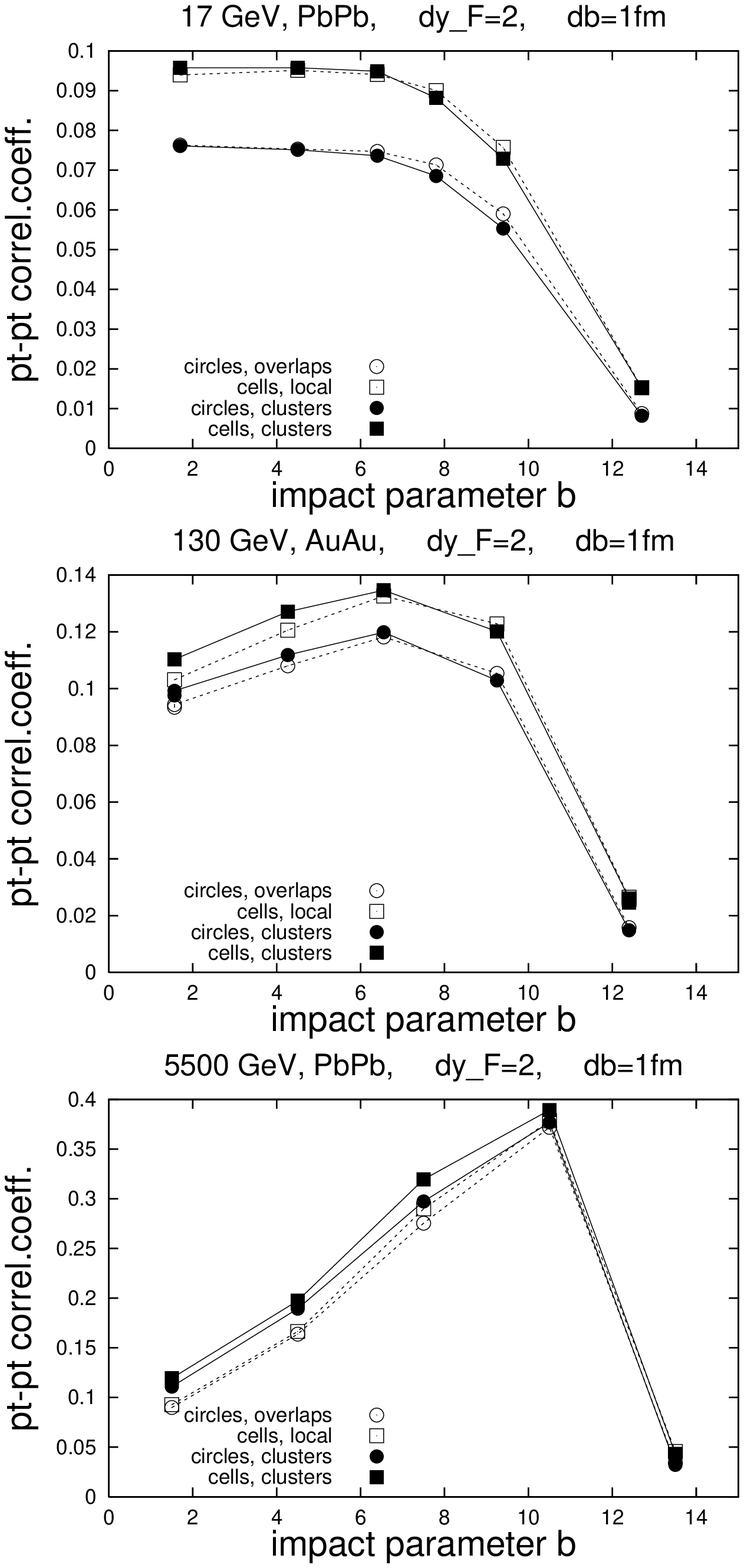}
 }
 \caption{The $b^{}_{p_t-p_t}$ correlation coefficient as a function of the impact parameter $b$
(at $db=1$ fm) for tree choices of the initial energy: $\sqrt{s}=17;\ 130$ and $5500$ GeV.
}
\end{figure}
%%%%%%%%%%%%%%%%%%%%%%%%%%%%%%%%%%%%%%%%
\section{Results of the calculations}
In Figs.1-3 the results of the MC calculations of these correlation coefficients
are presented for nucleus-nucleus collisions at different values of the centrality.
In all figures ($\circ$) and ($\bullet$) denote the results of calculations
in the framework of the original SFM (with the taking into account the real geometry
of merging strings) for its \textit{local (overlaps)} and \textit{global (clusters)}
versions correspondingly.
The ($\Box$) and (\rule[0.5mm]{2mm}{2mm}$\,$) denote the results of calculations
in the framework of the cellular analog of SFM
for its \textit{local} and \textit{global (clusters)}
versions. All presented results are
for the forward rapidity window of 2 unit length ($\DF=2$).
The lines are only to guide the eye.

In Fig.1 we present the $b^{}_{n-n}$ correlation coefficient
for AuAu collisions at RHIC energy and
in Fig.2 we present the $b^{}_{p_t-n}$ correlation coefficient
for PbPb collisions at LHC energy.
In both figures the calculations are fulfilled three times:

1) at fixed values of impact parameter ($db=0$),

2) with impact parameter fluctuations within 1 fm window ($db=1$),

3) with impact parameter fluctuations within the whole class of centrality ($db=class$)
  (for LHC by convention this value is taken to be equal 3 fm, $db=3$)
We see that the impact parameter fluctuations at a level of a few fermi
significantly change the magnitude of correlation coefficients.
Note also that all results obtained in the framework of the original SFM
and its cellular analog for their local and global versions practically
coincide,
except for the $p_t$-$n$ correlation at fixed value of impact parameter at LHC energy,
where the correlation coefficient $b^{}_{p_t-n}$ is very small.

In Fig.3 the energy dependence of the $b^{}_{p_t-p_t}$ correlation coefficient
is presented. The calculations are made for the 1 fm impact parameter window ($db=1$).
In this case we see the considerable rise of $p_t$--$p_t$ correlation
coefficient from SPS to LHC energies.

\section{Conclusion}

In the case with the real nucleon distribution density of colliding nuclei
the MC calculations of the long-range correlation functions
at different values of impact parameter are done.
For $n$--$n$ and $p_t$--$n$ correlations it is shown that
the impact parameter fluctuations at a level of a few fermi,
unavoidable in the experiment, significantly
change the magnitude of correlation coefficients
for all centrality classes as compared to ones
calculated earlier at the fixed values of impact parameter \cite{EPJC04}.

It is shown also, that for the $p_t$--$p_t$ correlation the event-by-event correlation
between event mean values of transverse momenta of the particles emitted
in two different rapidity intervals does not decrease to zero
with the increase of the number of strings
in contrast with the correlation between the transverse momenta
of single particles produced in these two rapidity windows
which was studied earlier \cite{EPJC04}.

The rise of $p_t$--$n$ and especially $p_t$--$p_t$ correlation
coefficients is found when one passes from SPS to RHIC and LHC
energies.

\subsection*{Acknowledgments}

The authors would like to thank M.A.~Braun
and G.A.~Feofilov for numerous valuable
encouraging discussions.
The work has been partially supported by
the Russian Foundation for Fundamental Research
under Grant No. 05-02-17399-a.

\end{document}